\documentclass[final,3p,times,twocolumn]{elsarticle}

\usepackage{amssymb}
\usepackage{color}
\usepackage{subfigure}
\usepackage{lineno}

\journal{NIM A}
\def\degree{${}^{\circ}$}
\begin{document}

\begin{frontmatter}



\title{Radiation effects on NDL prototype LGAD sensors after proton irradiation}
\author[label1,label3]{Yuhang Tan}
\author[label1,label3]{Tao Yang}
\author[label1,label3]{Suyu Xiao}
\author[label1,label3]{Kewei Wu}
\author[label5]{Lei Wang}
\author[label5]{Yaoqian Li}
\author[label5]{Zhenwei Liu}
\author[label1,label2]{Zhijun Liang}
\author[label4]{Dejun Han}
\author[label4]{Xingan Zhang}
\author[label1,label2]{Xin Shi\corref{cor1}}
\ead{shixin@ihep.ac.cn}

\address[label1]{Institute of High Energy Physics, Chinese Academy of Sciences, 19B Yuquan Road, Shijingshan District, Beijing 100049, China}
\address[label2]{State Key Laboratory of Particle Detection and Electronics, 19B Yuquan Road, Shijingshan District, Beijing 100049, China}
\address[label3]{University of Chinese Academy of Sciences, 19A Yuquan Road, Shijingshan District, Beijing 100049, China}
\address[label4]{Novel Device Laboratory, Beijing Normal University, No.19 Xinjiekouwai Street, Haidian District, Beijing 100875, China}
\address[label5]{China Institute of Atomic Energy, Beijing 102413, China}
\cortext[cor1]{Corresponding author}

\begin{abstract}
We study the radiation effects of the Low Gain Avalanche Detector (LGAD) sensors developed by the Institute of High Energy Physics (IHEP) and the Novel Device Laboratory (NDL) of Beijing Normal University in China. These new sensors have been irradiated at the China Institute of Atomic Energy (CIAE) using 100 MeV proton beam with five different fluences from 7$\times10^{14}$ $n_{eq}/cm^2$ up to 4.5$\times10^{15}$ $n_{eq}/cm^2$. The result shows the effective doping concentration in the gain layer decreases with the increase of irradiation fluence, as expected by the acceptor removal mechanism. By comparing data and model gives the acceptor removal coefficient $c_{A}$ = $(6.07\pm0.70)\times10^{-16}~cm^2$, which indicates the NDL sensor has fairly good radiation resistance.

\end{abstract}


% \begin{highlights}
% \end{highlights}


\end{frontmatter}

%\linenumbers

\section{Introduction}
To meet the high luminosity challenge for detectors in high energy physics (HEP), a new type of silicon technology - Low Gain Avalanche Detector (LGAD) is in development as 4 dimensions tracking detectors due to its good time resolution (less than 30 ps) \cite{Pellegrini:2014lki,Cartiglia:2015iua}. Currently, the main obstacle of the LGAD sensor is that the time resolution will degrade with the increase of irradiation fluence. When the irradiation fluence reaches 3$\times$$10^{15}$$n_{eq}/cm^2$, the time resolution is worse than 60 ps. Understanding the radiation effect is an essential step to mitigate the deterioration of the LGAD performance.\par

\begin{figure}[ht]
\center
\begin{minipage}[ht]{0.6\linewidth}
\includegraphics[width=1.0\textwidth]{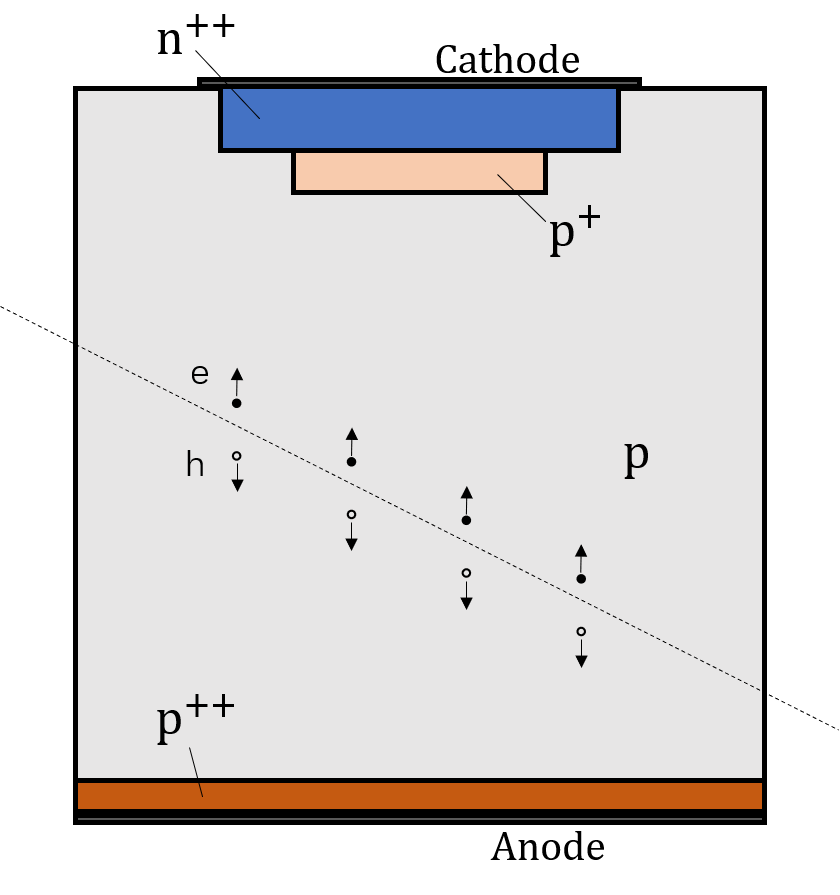}
\end{minipage}
\caption{Schematic view of the LGAD concept.
\label{fig:sensor}}
\end{figure}

The structure of the LGAD sensor is $n^{++}-p^{+}-p-p^{++}$ as shown in Fig.\ref{fig:sensor}, where the high electric field near the gain layer ($p^{+}$ layer) can achieve impact ionization \cite{Pellegrini:2014lki}. Meanwhile, the low gain (5 $\sim$ 50) characteristic of the LGAD sensor can multiply the signal without multiplying the noise, so the signal-to-noise ratio and time resolution are improved. The good timing performance of the LGAD sensor is based on the gain factor, which depends on it's effective doping concentration and profile. According to the acceptor removal mechanism \cite{Ugobono:2018tla}, the radiation reduces the effective doping concentration in the gain layer, thus deteriorates the time resolution. \par
\par
The LGAD has been pioneered by the Centro Nacional de Microelectronia (CNM) Barcelona \cite{Pellegrini:2014lki} in close collaboration with the RD50 collaboration \cite{Mazza:2667164}. The Institute of High Energy Physics (IHEP) of Chinese Academic of Sciences joined the RD50 and ATLAS High-Granularity Timing Detector (HGTD) collaboration to develop LGAD sensor with the Novel Device Laboratory (NDL) of Beijing Normal University  in China. The measurement of basic characteristics \cite{Yang:2019fzk} and test beam of first prototype IHEP-NDL sensor before irradiation \cite{1796576} have been completed. It's essential to evaluate the performance of NDL sensor after irradiation and the related irradiation effects.

To investigate the radiation effect, the NDL sensors were irradiated at the China Institute of Atomic Energy (CIAE) using 100 MeV proton at five fluence points: 7$\times10^{14}$, 1$\times10^{15}$, 2$\times10^{15}$, 3$\times10^{15}$ and 4.5$\times10^{15}$ ($n_{eq}/cm^2$), where the proton irradiation fluence has converted to 1 MeV neutron equivalent fluence with the conversion factor 1.276 \cite{Allport:2019kvs}. The irradiation induced macroscopic effect can be observed as follows: the changing on the depletion voltage, the decrease of the effective doping concentration in the gain layer and the loss of the gain factor. The study of these macroscopic effects of LGAD with different fluences will shed light on the microscopic understanding of the acceptor removal mechanism\par
The contents of this paper are organized as follows. The equipments and processes of the irradiation experiment are described in Section 2. The performance of the irradiated sensor are shown in Section 3, followed by a conclusion section.\par
\section{The NDL sensor and irradiation experiments}
\subsection{The NDL sensor}
\begin{figure}[h]
\center
\begin{minipage}[ht]{0.6\linewidth}
\includegraphics[width=1.0\textwidth]{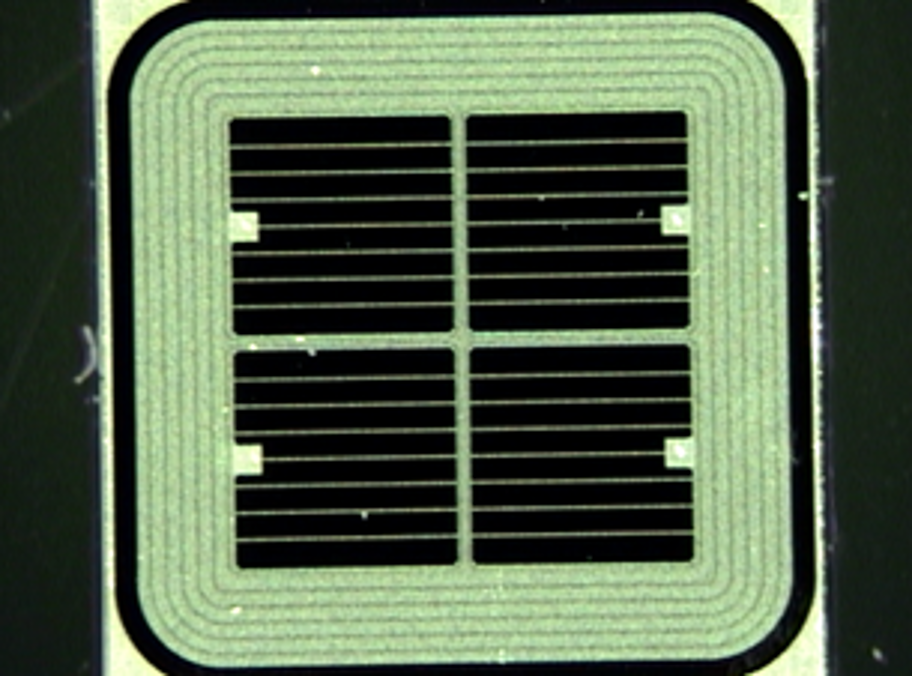}
\end{minipage}
\caption{The image of the NDL-BV60 sensor.
\label{fig:NDL_sensor}}
\end{figure}

This paper focuses on one type of NDL sensor BV60, with image shown in Fig.\ref{fig:NDL_sensor}. The front side of one sensor has four square 1.00$\times$1.00 mm$^{2}$ pads with backside fully metalized. Each pad has a square readout electrode, and surrounded by six floating guard rings. The active thickness is 33 $\mu$m processed on epitaxy silicon ($\sim$300$\Omega$$\cdot$cm) with Boron doped $p^{+}$ layer.

\subsection{Experiment setup}
\begin{figure}[ht]
\begin{minipage}[ht]{0.49\linewidth}
\includegraphics[width=1.0\textwidth]{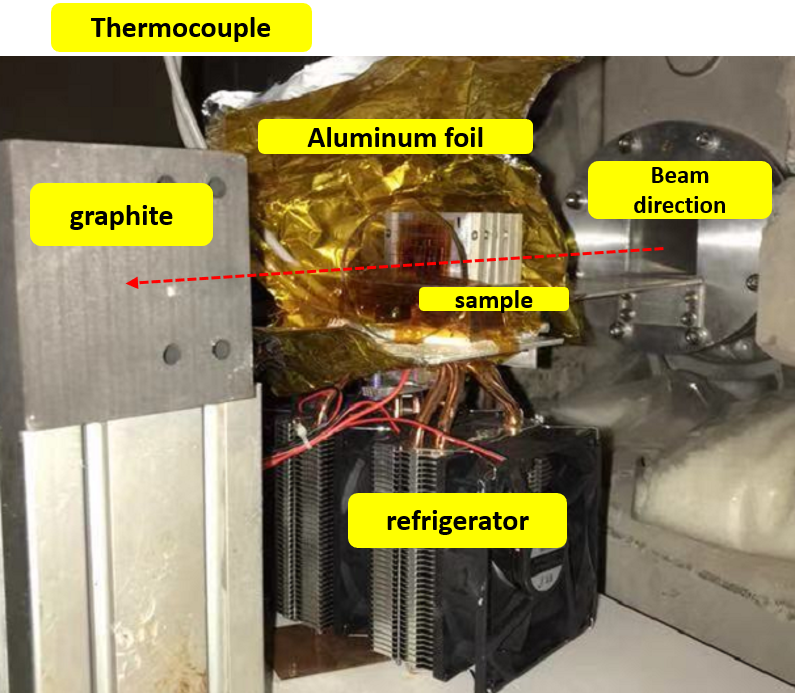}
\end{minipage}
\hfill
\begin{minipage}[ht]{0.49\linewidth}
\includegraphics[width=1.0\textwidth]{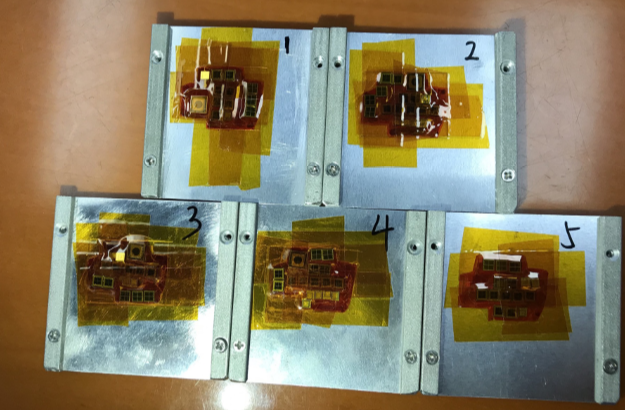}
\end{minipage}
\hfill
    \caption{The experiment setup at the proton beam line (left). LGAD sensors fixed on five aluminum plates with kapton tape for different radiation fluences (right).
    \label{fig:experiment}}
\end{figure}

Fig.\ref{fig:experiment} (left) shows the experiment setup including thermocouple, graphite, refrigerator, aluminum foil and sensors.  The beam direction is from right to left as indicated by the red line (color online). The beam uniformity is less than 1\% and the beam size is around 2$\times$2 cm$^{2}$, confirmed by simulation and fluorescent film test. During the experiment, the stability of the beam luminosity is around 1\% monitored every 30 minutes. The beam current is 100 nA corresponding to a beam flux of 10$^{11}$ hadrons/(cm$^{2}\cdot$s) converted from the ratio between current and elementary charge. The irradiation fluence is calculated by beam flux times the beam time. \par
The LGAD sensors are fixed with kapton tape on five aluminum plates for different fluences (Fig.\ref{fig:experiment} right). The five aluminum plates are arranged in sequence perpendicular to the beam direction. Based on the simulation, the attenuation of the beam over these plates is negligible. The irradiation fluence on each plate is rather uniform. \par
During the whole irradiation process, the temperature of the sensors was kept below 0\degree C with a Peltier and monitored by thermocouple to prevent the thermal annealing, which could alter the effective doping concentration and defects concentration \cite{hara2016charge}. The graphite block placed at the end of the beam serves as an absorber to prevent protons from hitting the interior wall of the terminal that can produce harmful substances. The sensors were annealed for 70 minutes at 60 \degree C after irradiation and then kept in fridge to prevent further annealing. \par

\section{The performance of irradiated sensor}
\subsection{Leakage current}
\begin{figure}[ht]
\begin{minipage}[t]{1.0\linewidth}
\includegraphics[width=1.0\textwidth]{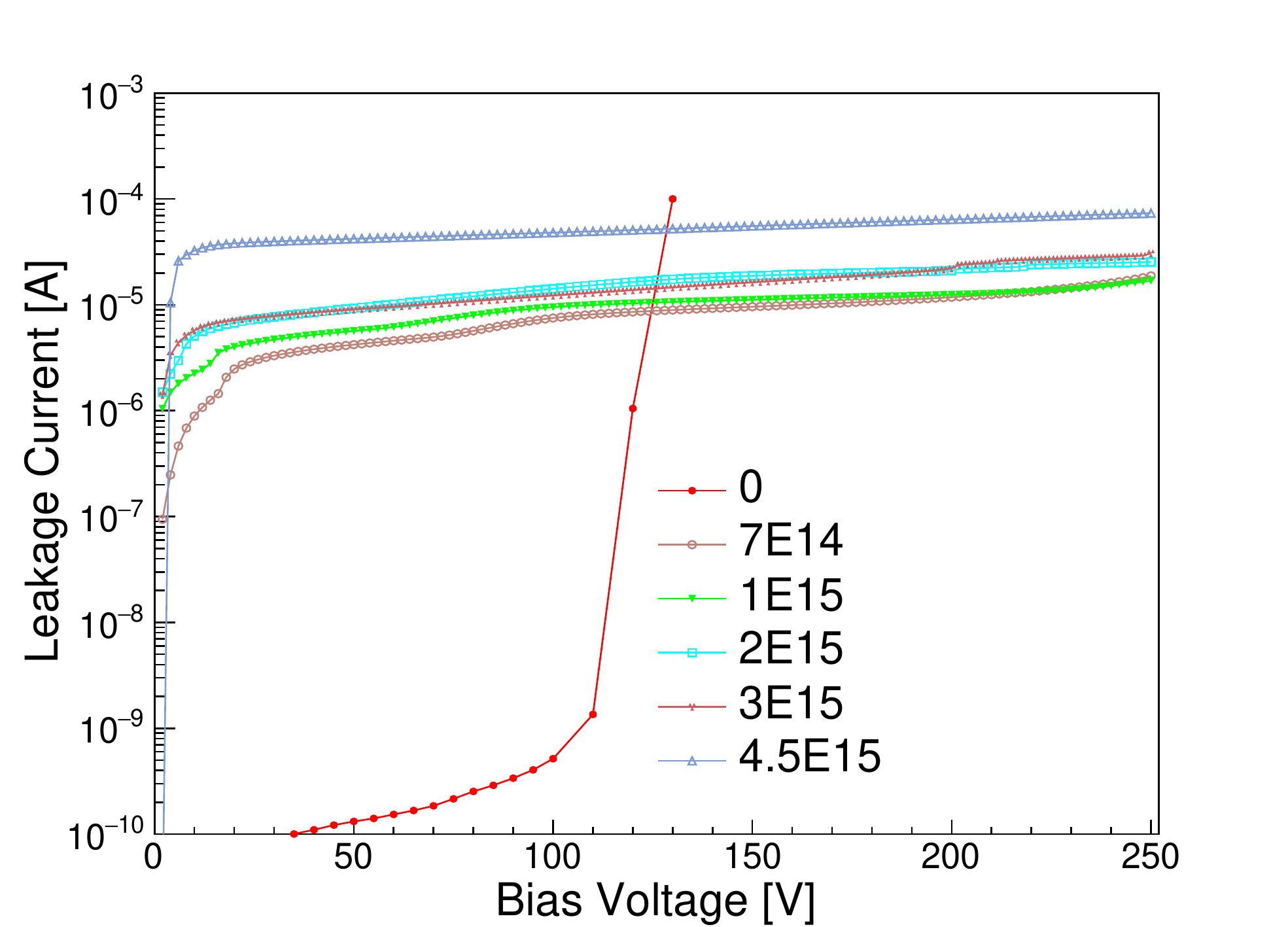}
\end{minipage}
    \caption{Leakage current for NDL sensor measured at different fluences from 0 to 4.5 $\times10^{15}$ ($n_{eq}/cm^2$) at room temperature. The leakage current increases nonlinearly with the fluence.
    \label{fig:IV}}
\end{figure}

Usually, the leakage current of a standard silicon sensor increases linearly with irradiation fluence \cite{Moll:1999nh}. However, the situation of LGAD is more complicated due to the gain layer. The leakage current of NDL BV60 increased by 5 to 6 orders of magnitude after irradiation and has non-linear features at different fluences as shown in Fig.\ref{fig:IV}. For LGAD, the leakage current depends on the generation current $I_{gen}$ and the gain factor M: $I_{leak}$=M$\cdot I_{gen}$ \cite{Kramberger:2018cun}. Usually $I_{gen}$ in silicon sensor follows \cite{Moll:1999nh}:
\begin{equation}
\label{E2}
I_{gen}=\alpha\cdot S\cdot d\cdot\Phi_{eq}
\end{equation}
where $\alpha$ is the irradiation damage rate, S is the active area, d is the thickness and $\Phi_{eq}$ is the irradiation fluence. $\alpha$ is about 3.9$\times10^{-17}$ A/cm \cite{Moll:1999nh} with annealing time of 70 min at 60\degree C. Since S and d are constant for a given sensor, the increase of irradiation fluence leads to the increase of $I_{gen}$. M decrease with fluence because radiation reduces the effective doping concentration of gain layer. As the gain decreases and the $I_{gen}$ increases with irradiation, the $I_{leak}$ does not necessarily increase monotonically with fluence. For example, the $I_{leak}$ of sensors at 2$\times10^{15}$ $n_{eq}/cm^2$ and 3$\times10^{15}$ $n_{eq}/cm^2$ are equal. \par
The gain factor M can be estimated from the measured $I_{leak}$ and the calculated $I_{gen}$. For five irradiation fluences from 7$\times10^{14}$ $n_{eq}/cm^2$ up to 4.5$\times10^{15}$ $n_{eq}/cm^2$, the M are 4.7, 4.4, 3.6, 2.3, and 11.0 at the bias voltage of 50 V. The sudden increase of M at the highest fluence 4.5$\times10^{15}$ $n_{eq}/cm^2$ in some NDL sensors might caused by unknown mechanism that leads to the miscalculation of $I_{gen}$. Further tests with $\beta$ source is necessary in order to justify the calculation procedure.
\subsection{C-V characteristics}
\begin{figure}[ht]
\begin{minipage}[t]{1.0\linewidth}
\includegraphics[width=1.0\textwidth]{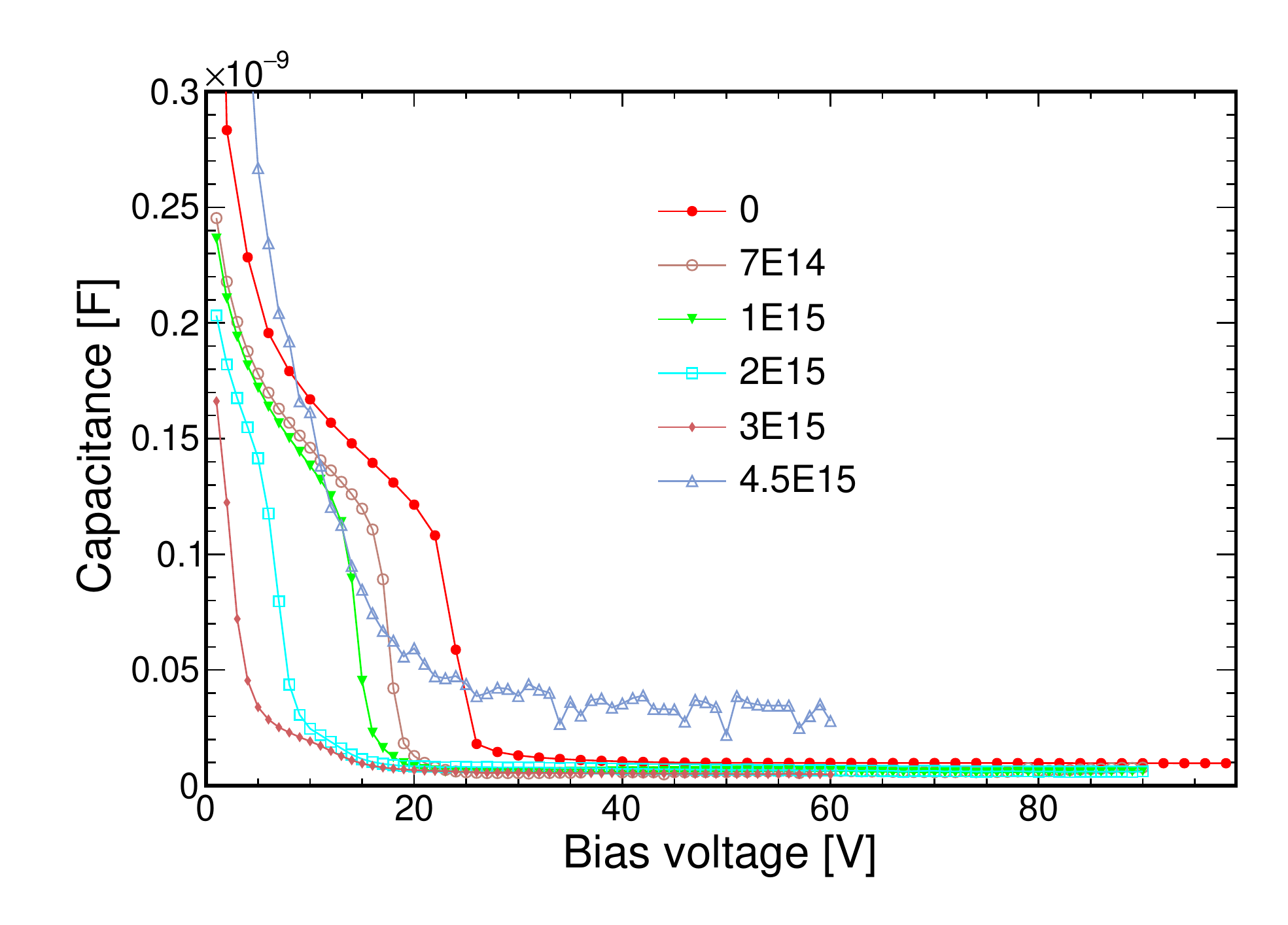}
\end{minipage}
    \caption{ C-V scan of the NDL sensor measured at different fluences given in [$cm^{-2}$] at room temperature.
    \label{fig:CV}}
\end{figure}

The capacitance-voltage (C-V) scans are used to evaluate the full depletion voltage ($V_{FD}$). The measured C-V curves for NDL sensors before and after irradiation are shown in Fig.\ref{fig:CV}. The curves exhibit a sharp fall-off at the bias voltage ($V_{GL}$) where the gain layer is fully depleted. As the irradiation fluence increase, the decrease of $V_{GL}$ indicates the reduction of the effective doping concentration in gain layer which agrees with the acceptor removal mechanism. After the full depletion of gain layer, the capacitance quickly reaches saturation due to lightly doped bulk. \par
\begin{figure}[ht]
\begin{minipage}[t]{1.0\linewidth}
\includegraphics[width=1.0\textwidth]{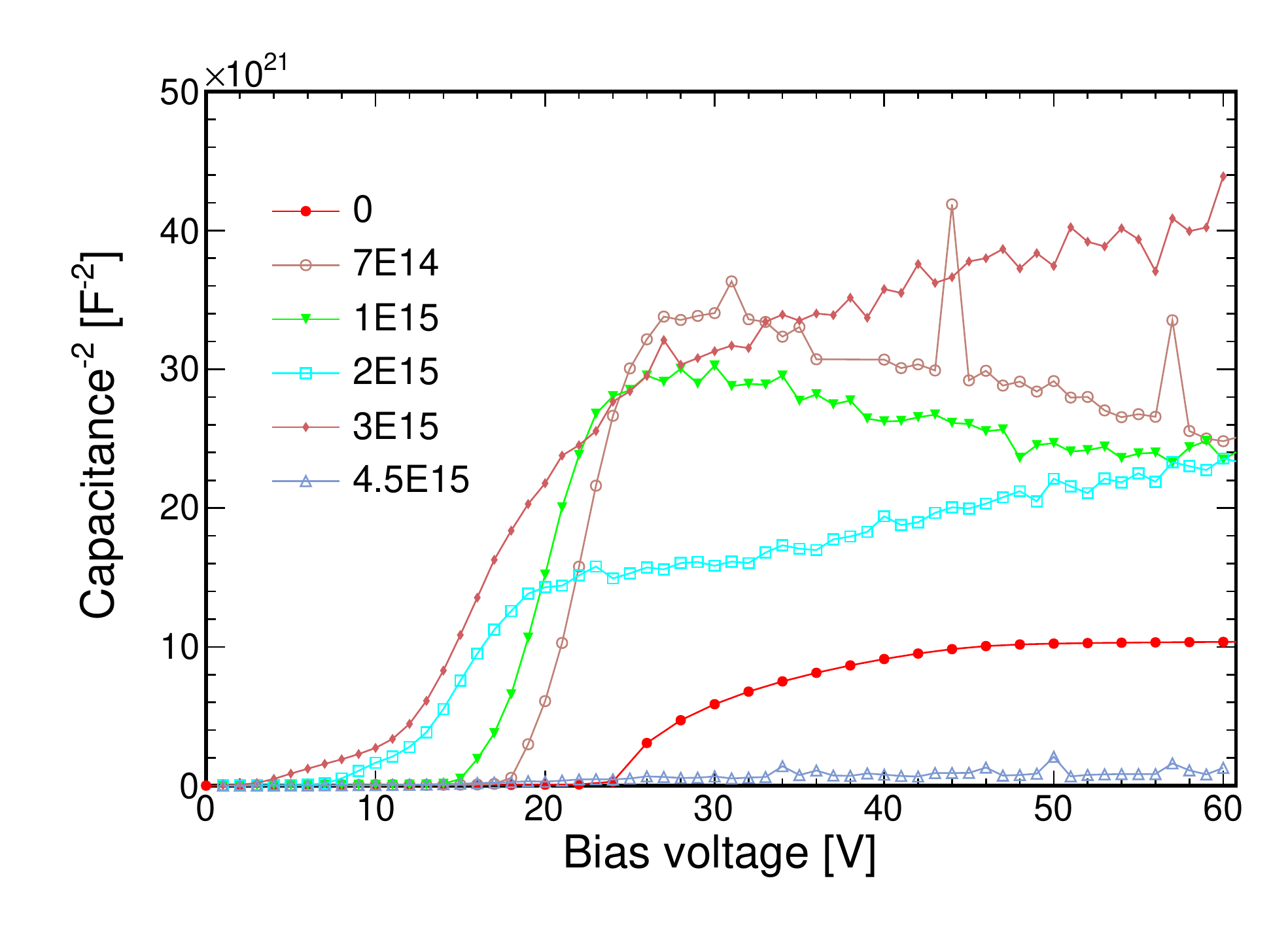}
\end{minipage}
    \caption{ The $1/C^{2}$ distribution of the NDL sensor measured at different fluences given in [$cm^{-2}$] at room temperature.
    \label{fig:C2V}}
\end{figure}

For the $1/C^{2}$-V curves are shown in Fig.\ref{fig:C2V}, the $V_{GL}$ is recognized as the point where the curves start a sharp increase and $V_{FD}$ is the point where the curves turn to flat after sharp increase. The difference between $V_{FD}$ and $V_{GL}$ is proportional to the doping concentration of the sensor bulk: $V_{bulk}$= $V_{FD}$ - $V_{GL}$. The radiation effect of acceptor removal in the gain layer is clearly visible by the decrease of $V_{GL}$. Due to the change of $V_{bulk}$ is irregular with fluence, it can not be justified that the radiation can generate acceptor-like defects in bulk layer. There is no clear explanation of the larger $V_{FD}$ and capacitance at 4.5$\times10^{15}$ $n_{eq}/cm^2$. Further low-temperature research and more data are necessary.\par
\subsection{Acceptor removal in NDL sensor}
\begin{figure}[ht]
\begin{minipage}[t]{1.0\linewidth}
\includegraphics[width=1.0\textwidth]{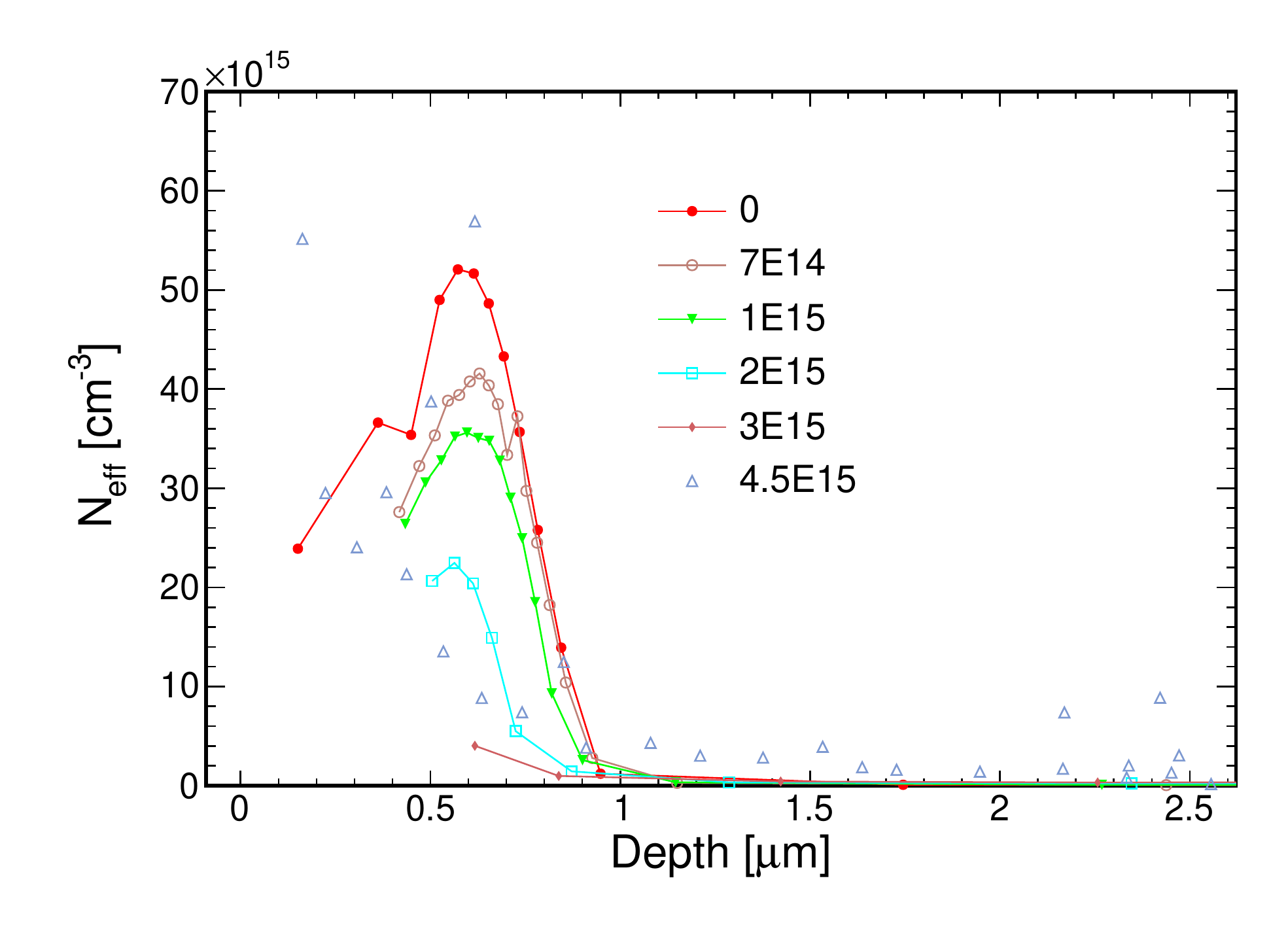}
\end{minipage}
    \caption{Effective doping concentration ($N_{eff}$) dependence on doping depth ($W_{dep}$) for the NDL sensors measured at different fluences given in [$cm^{-2}$] at room temperature.
    \label{fig:edc}}
\end{figure}

On the macroscopic level, the acceptor removal effect in LGAD can be observed as the radiation induced decrease of depletion voltage ($V_{dep}$), respectively effective doping concentration ($N_{eff}$). The experimental data available are based on the measurement of depletion depth ($W_{dep}$), which is converted under the assumption of a homogeneous space charge into the $N_{eff}$ \cite{Moll_lgad_acceptor}:
\begin{equation}
\label{E5}
N_{eff}= \frac{2V_{bias}\epsilon}{W_{dep}^{2}q_{0}}
\end{equation}
where $q_{0}$ is the elementary charge and $\epsilon$ the permittivity of silicon. $N_{eff}$ dependence on $W_{dep}$ for the NDL sensors measured at different fluences is shown in Fig.\ref{fig:edc}. The effective doping concentration $N_{eff}$ in the gain layer decrease with irradiation fluence, which agrees well with the acceptor removal mechanism.
\par
\begin{figure}[ht]
\begin{minipage}[t]{1.0\linewidth}
\includegraphics[width=1.0\textwidth]{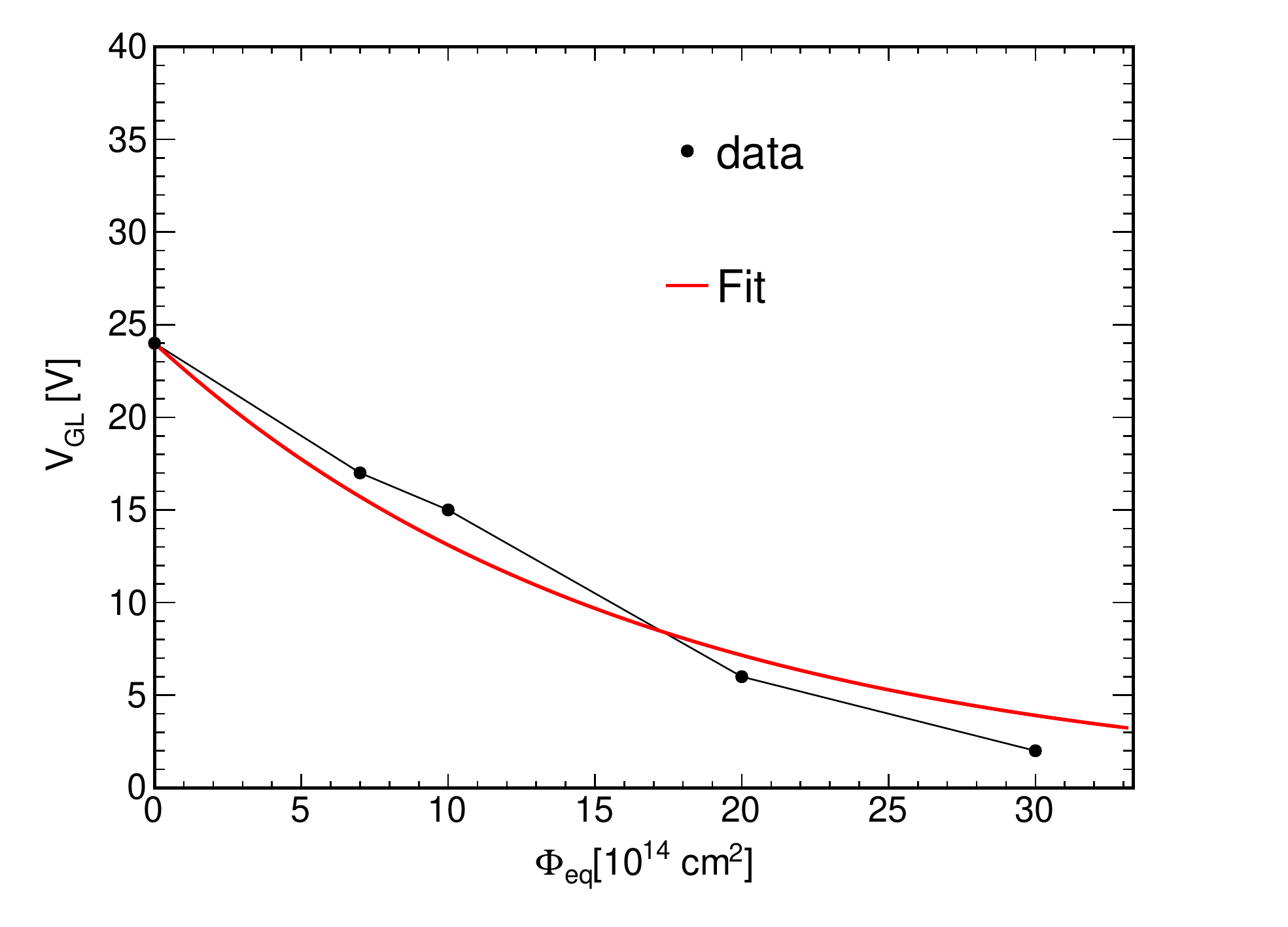}
\end{minipage}
    \caption{The $V_{GL}$ as function of irradiation fluences for the NDL sensor measured at room temperature.  The curve is fit to the data resulting in the acceptor removal coefficient $c_{A}$ = $(6.07\pm0.70)\times10^{-16}~cm^2$.
    \label{fig:tvmr}}
\end{figure}
To explain the acceptor removal effect at the microscopic level, the dopant Boron in the gain layer is removed from the substitutional lattice site and bound into defect complexes like $B_{i}O_{i}$ which no longer exhibit shallow acceptor properties leading to the decrease of $N_{eff}$ \cite{Moll_lgad_acceptor}. At the same time, radiation can generate acceptor-like defects so that $N_{eff}$ will increase with fluence. The combined effects are described by \cite{ferrero2019radiation}:
\begin{equation}
\label{E1}
N_{eff}(\Phi_{eq})=N_{A_0}e^{-c_{A}\Phi_{eq}}+g_{eff}\cdot\Phi_{eq}
\end{equation}
where $N_{A_0}$ is the acceptor's initial concentration before irradiation and $g_{eff}$ is the introduction rate of stable deep acceptors for irradiation with protons in this paper. Moreover, $N_{A_0}e^{-c_{A}\Phi_{eq}}$ is the acceptor removal term and $g_{eff}\Phi_{eq}$ is related to the acceptor-like defects. The model also explains the phenomenon that the $N_{eff}$ decrease first and then increase  in the bulk layer with fluence \cite{Moll:1999kv}. \par

Assuming the $N_{eff}$ in bulk layer is constant, $V_{GL}$ is proportional to an average concentration of Boron and the evolution of $V_{GL}$ with the irradiation fluences ~\cite{Kramberger:2015cga} can be described as:
\begin{equation}
\label{E6}
V_{GL}(\Phi_{eq})\approx V_{GL_0}e^{-c_{A}\Phi_{eq}}
\end{equation}
where $V_{GL_0}$ is the depletion voltage of the gain layer of the unirradiated sensor and $\Phi_{eq}$ is the irradiation fluence. By fitting the data with Eq.\ref{E6}, one can extract the acceptor removal coefficient $c_{A}$ as $(6.07\pm0.70)\times10^{-16}~cm^2$. The data is consistent with the acceptor removal model on macroscopic level. \par
\begin{figure}[ht]
\begin{minipage}[t]{1.0\linewidth}
\includegraphics[width=1.0\textwidth]{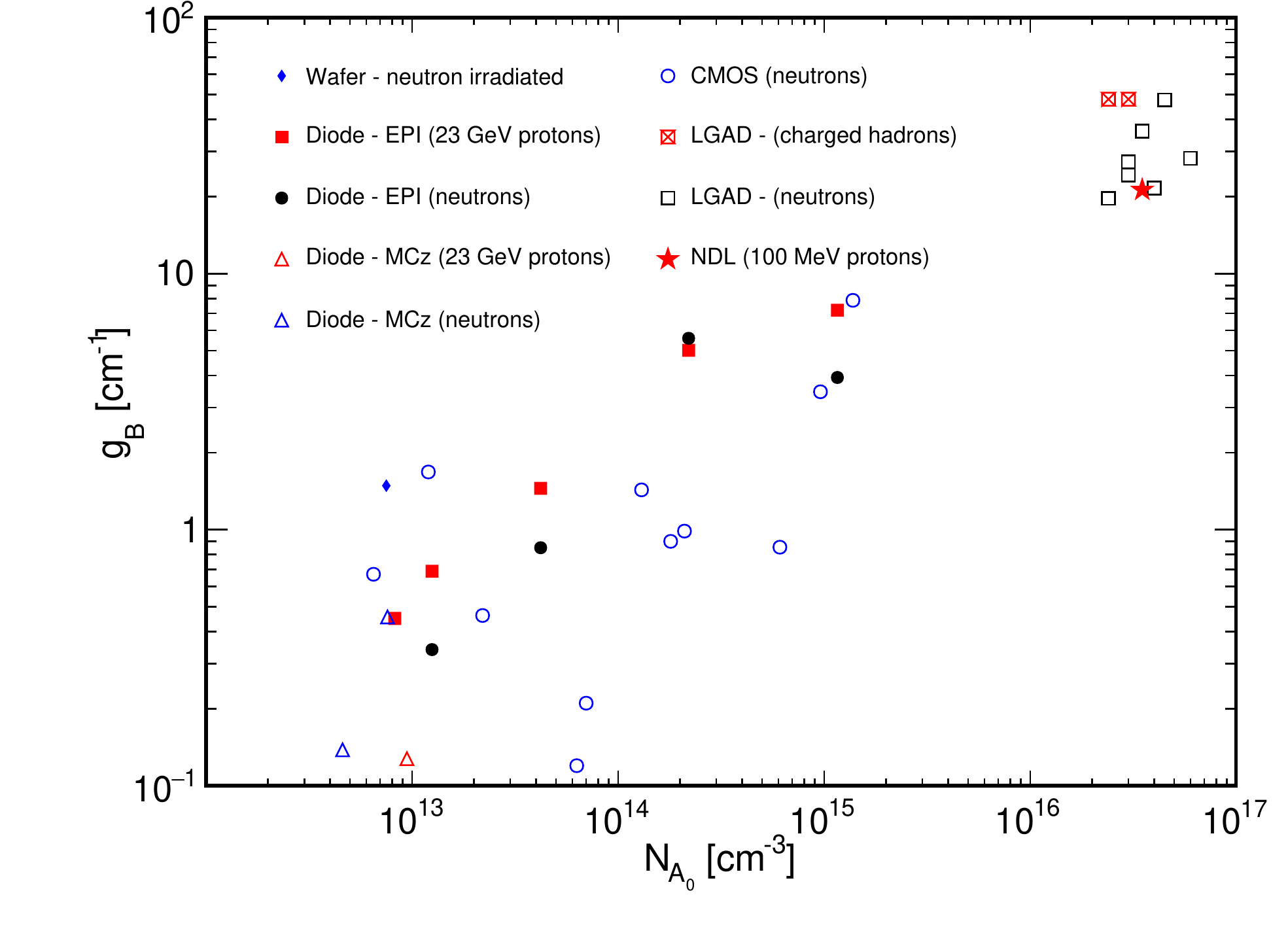}
\end{minipage}
    \caption{ (color online) The initial acceptor removal rate $g_{B}$ as function of the initial doping concentration for p-type silicon sensors \cite{Moll_lgad_acceptor}. The red star is the result of this experiment.
    \label{fig:irradiation_gb}}
\end{figure}

The parameter $g_{B} = c_{A} \times N_{A_0}$ represents the number of deactivated acceptors per unit volume and fluence. Fig.\ref{fig:irradiation_gb} shows the initial acceptor removal rate $g_{B}$ as function of the initial doping concentration for p-type silicon sensors. The red star is the result of this experiment, which is the agreement with the results of other LGAD experiments indicating the NDL sensor has fairly good radiation resistance. However, there is still no reasonable explanation that the $g_{B}$ of sensors with high initial doping concentration is much larger than low initial doping concentration, so it is essential to clarify and parameterize the drastic changes of $g_{B}$. More microscopic data of the irradiation LGAD sensor is needed to better understand the acceptor removal mechanism. \par

\section{Conclusion}
The irradiation on the LGAD sensor will degrade its performance in terms of gain and timing resolution. To study the radiation effects, the first irradiation experiment at CIAE for the NDL sensor was completed using the 100 MeV proton beam with five different fluences from 7$\times10^{14}$ $n_{eq}/cm^2$ up to 4.5$\times10^{15}$ $n_{eq}/cm^2$. The result showed the radiation effect on the leakage current, capacitance and the effective doping concentration of LGAD. The effective doping concentration decreases with the increase of irradiation fluence which can be explained by the removal of dopant Boron in the gain layer from the substitutional lattice site and bound into defect complexes $B_{i}O_{i}$. The acceptor removal model fits the data well to give the acceptor removal coefficient $c_{A}$ as $(6.07\pm0.70)\times10^{-16}~cm^2$. \par
 Comparing with the radiation resistance of other LGAD sensors, for example the acceptor removal coefficient $c_{A}$ is about 4.70 $\times10^{-16}$ $cm^2$ for FBK sensor and 8.10 $\times10^{-16}$ $cm^2$ for CNM sensor \cite{Moll_lgad_acceptor}, the NDL sensor shows fairly good radiation resistance. While the consistency between data and acceptor removal model is good in the gain layer, a comprehensive understanding of the microscopic defect formation level is still lacking. \par
 Nowadays, some dedicated techniques, such as deep-level transient spectroscopy (DLTS) and thermally stimulated current (TSC)) have been used for the detection and characterization of all radiation induced defects in standard silicon, but there are still lacking the results of defects on LGAD. Analysis of electrically active defects and the radiation induced changes in the electrical characteristics of LGAD will help to connect the phenomenon of acceptor removal on LGAD and the defect formation measurement in the future. %This may be a feasible direction for the theory of defects and defect kinetics modeling.\par

\section*{Acknowledgment}
We acknowledge the help from the CERN RD50 collaboration and ATLAS HGTD collaboration. This work was supported by the National Natural Science Foundation of China (No. 11961141014), the State Key Laboratory of Particle Detection and Electronics (SKLPDE-ZZ-202001), and a funding of the Chinese Academy of Sciences (Y6291150K2). We also want to thank Michael Moll for providing data for Fig.\ref{fig:irradiation_gb} and many useful discussions.

%We would thank the Novel Device Laboratory for providing sensors and the staff of the China Institute of Atomic Energy for the irradiation equipment. 
 
\bibliographystyle{unsrt}
\bibliography{p3_lgad_ndl_ciae}

\end{document}